\documentclass{article}
\usepackage[utf8]{inputenc}
\pdfoutput=1
\usepackage{fullpage, multicol, verbatim, enumitem, changepage, mdframed, outlines,caption}
\usepackage{xcolor}
\usepackage{xparse}
\usepackage[most]{tcolorbox}

\usepackage{xparse}
\usepackage{lipsum}
\usepackage{scrextend}
\usepackage{subfiles}
\usepackage{nameref}
\usepackage{listings}
\usepackage{wrapfig}
\usepackage{appendix}
\usepackage{pdfpages}

\usepackage{authblk}

\title{How Well do Remote Labs Work?\\A Case Study at Princeton University}

\author[1]{Saumya Shivam}
\author[1]{Kasey Wagoner}
\affil[1]{Joseph Henry Laboratories of Physics, Jadwin Hall, Princeton University, Princeton, NJ, USA 08544}

\begin{document}
\maketitle

\section{Introduction}
The onset of the COVID-19 pandemic forced many universities to move to virtual instruction during the spring 2020 semester. The transition to remote learning was abrupt and overwhelming for teachers of all subjects, all across the US. Nowhere was this more true than in science lab courses. The experience nevertheless provides an opportunity to investigate the optimal design of remote labs, with similar learning goals as in-person labs. In this study we explore the three most common approaches to remote labs: recorded experiments, applet-based experiments, and at-home projects. We use surveys and interviews to make two comparisons: remote labs vs. in-person labs; the different types of remote labs. Examining these two questions we find that remote labs perform as well as in-person labs and students learn the most from at home physics experiments while also enjoying those the most. 

We first describe the structure of the labs, and the different types of remote labs, then describe the surveys and interviews we use to study the efficacy of such labs, and finally analyze the results from the surveys.

\section{Our Labs}
This study was conducted in Physics 101 at Princeton University. In the spring of 2020 the course had 62 students, 40\% sophomores, 50\% juniors, and 10\% seniors. To first order, this course is a typical introductory algebra-based, course for non-majors.  The pedagogical structure follows the Investigative Science Learning Environment (ISLE) model \cite{Etkina2015} and when on campus, the labs are typical ISLE labs \cite{Visnjic2015}. The primary goal of our labs is for students to develop an understanding of the process of experimental physics, not to reinforce concepts.

\subsection{Remote Labs}
Princeton University transitioned to remote instruction at the midway point of the spring 2020 semester. After the transition, students completed three labs: one recorded experiment, one project experiment, and one virtual experiment; each of these are described below. The primary motivation for assigning different types of labs was to create a situation where we could assess the efficacy of the different approaches. 

The first remote lab that the students completed involved a set of recorded experiments. The experiments were exactly what the students would have done, had they been on campus. The lab covers magnetic induction with ``observation'' and ``testing'' experiments which are typical of the ISLE pedagogy. The lab consisted of three videos of instructors performing experiments. After watching the videos, the students worked through a worksheet in which they were supposed to use observations to develop a model of magnetic induction; to this point they had not encountered the magnetic induction in any part of the course. 

The second remote experiment was a three-week long, student-defined project. Students were given three different project options. This was done to ensure that all students would be able to complete the project, regardless of their circumstances or ability to access tools (e.g. internet, equipment, etc.); each of those options is described below.  For each of these options, students were given examples projects to learn from and detailed rubrics that described exactly how their projects would be graded. The only restriction on the content of the project was that it should come from material covered that semester (E\&M, optics, circuits, atomic physics, nuclear physics). 
\begin{itemize}
\item Virtual Experiment: For this type of project students were tasked with using a web applet to ``experimentally'' answer a question which they had defined themselves (e.g. PhET \cite{PhET2008}; students were given a spreadsheet with links to a large number of applets). They were required to design an experimental method and perform the measurements using the applet. They then used standard methods to analyze the data and answer the question. 
\item At Home Physics: For this type of project students were tasked with using materials available to them to experimentally answer a question which they had defined themselves. They were required design an experimental method and perform the measurements. They then used standard methods to analyze the data and answer the question. To aid them, they were provided with suggestions software that make use of the numerous sensors in their smart devices. 
\item Algodoo Project: For this project students were tasked with creating a new ``scene'' in Algodoo (this is a software that has nearly all of physics built into it: www.algodo.com). They had complete freedom to make the scene into whatever they wanted. To give some inspiration, they were given suggestions of creating a video game and creating a model of the human eye. To aid them, they were given a number of example scenes which the instructors had previously created \cite{WashULabs2015}. 
\end{itemize}

The third remote lab was a one-week long set of virtual experiments. For this lab, students were tasked with making two different instructor-defined “measurements” using two PhETs (Alpha Decay and Beta Decay). For this, they were required to take “data”, analyze the data, and submit their analyzed data. The statistical nature of radioactive processes allowed us to probe the statistical uncertainties, how to mitigate them and how to display data with statistical fluctuations, giving the lab the feeling of a real experiment.

\subsection{Instructor Interactions}
The students in the class were divided into three sections of $\sim$ 20 students each. Within each section the students were recommended to form a group of three (occasionally two) at the beginning of the semester. During the first half of the semester, with in-person labs, two AIs (Assistant Instructors, who are typically graduate students in the Physics department) were present for each lab session. The AIs typically summarized the concepts and equipment function at the beginning of each lab, and assisted the students during the lab as needed.

After the transition to remote instruction, the nature of interactions between students and the AIs changed as well. For the recorded experiment, the AIs were available to answer questions about the lab through email. For the project lab, each AI was assigned a section, with the idea that the same AI would guide a given lab group for the three weeks of the project. During the first week of the project the aim was to help the students in choosing a well defined question that they would begin to answer. In this process they would learn about the different types of applets available online, think about the equipment and setup that could be used for a home experiment, and become familiar working with Algodoo. At the same time, the students were expected to relate their experiments with the content of the course covered up to that point (including electrostatics, DC circuits, magnetic induction and optics). The second week was typically about consolidating the experiment and answering questions about analysing any initial data, and the final week was supposed to help the students put everything together and resolve any errors or discrepancies. For each of these three weeks, the AIs scheduled a virtual meeting with each lab group, keeping track of their progress and providing guidance whenever required. In addition to the AIs, the instructor was also available for discussions during the office hours for the course. For the third remote experiment, a virtual experiment on radioactivity, each lab group was provided the option to sign up for a virtual meeting with an AI where they could ask questions or seek clarifications. 

Both before moving to remote instruction and after, each lab group was required to submit a report summarizing their experiment, with the students expected to follow the standard practices with aspects like creating plots and calculating uncertainties in the result. This report was then graded by the AIs, using ISLE rubrics ~\cite{Etkina2015}. New rubrics were developed for the project experiment. These rubrics were inspired by the ISLE rubrics, but were expanded to include topics such as creativity. The full set of rubrics for the project lab can be seen in Appendix \ref{app:Rubrics}. 

\section{Analysis}

\subsection{Measurement Instruments}
In order to assess the efficacy of our off-campus labs, we employed a number of different instruments. We gave students three different surveys and did a series of interviews. 

\subsubsection{Surveys}
Our students completed two different surveys from which we can draw information about the remote labs. The first was the E-CLASS survey developed by Zwickl et. al \cite{ECLASS2014}. The second was an optional, anonymous one created by the course instructors to get feedback on the transition to remote instruction. \\

\noindent \textbf{E-CLASS}: The E-CLASS survey series assesses how students' responses to a number of questions compare to the responses of experts. It is a pre/post series, so it can assess how students' alignment with experts shifted during the time between the pre and post surveys. Our students completed the series two different times; it was the same pool of students taking both survey series. The series was first taken during the fall semester, when all labs were in-person (69 students, 58\% of students responded); the pre was taken before any labs, and the post was taken after all labs were complete. The series was taken the second time during the spring, when we transitioned to remote learning (43 students, 69\% of students responded); the pre was taken mid-semester before any remote labs, and the post was taken after all remote labs were complete. This approach allowed us to compare E-CLASS results for in-person labs, to results for remote labs.\\ 

\noindent \textbf{Instructor Created Survey}: The second survey was and end-of-the-semester evaluation created by the instructors specifically for getting anonymous feedback on the transition online (completed by 15 students, 24\% of full enrollment). In this survey there were a few lab-specific questions. Each of these surveys gave us information to make comparisons: (i) in-person labs vs. remote labs; (ii) different types remote labs. 

\subsubsection{Interviews}
After the semester was complete, we conducted Zoom interviews with eight students (13\% of full enrollment). During the interview, students were asked only questions specific to the labs. Appendix \ref{app:Interview} has a full list of questions. 

The demographics of the interviewees closely followed those of the full course enrollment: 4 completed At Home Physics (AHP) projects, 4 completed Virtual Experiment (VE) projects (full enrollment: 43.5\% AHP, 50\% VE projects, 6.5\% Algodoo projects); course grades 5 As, 3 Bs (full enrollment 66\% As, 32\% Bs); 3 women, 5 men (full enrollment: women 63\%, men= 37\%). 

\subsection{Comparing in-person labs to remote labs}

\begin{figure}[h]
\centering
\includegraphics[height=0.2\textheight]{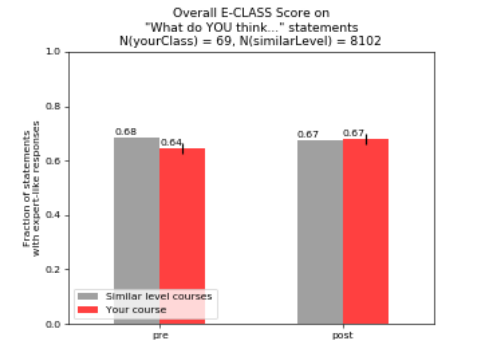}\hspace{5mm}\includegraphics[height=0.2\textheight]{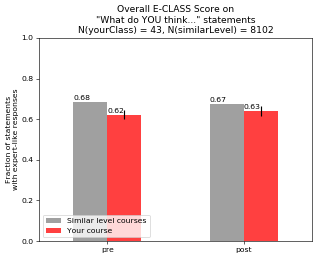}
\caption{Shown is a comparison of how expert-like student responses were to all of the questions on the E-CLASS. The fall semester (completely in-person) is shown on the left, while the scores on the right are for the spring semester, with remote instruction for part of the semester. Despite the sudden shift to remote labs, the overall score doesn't significantly decrease for the spring semester.}
\label{fig:overallcompare}
\end{figure}

\begin{figure}[h]
\centering
\includegraphics[height=0.25\textheight]{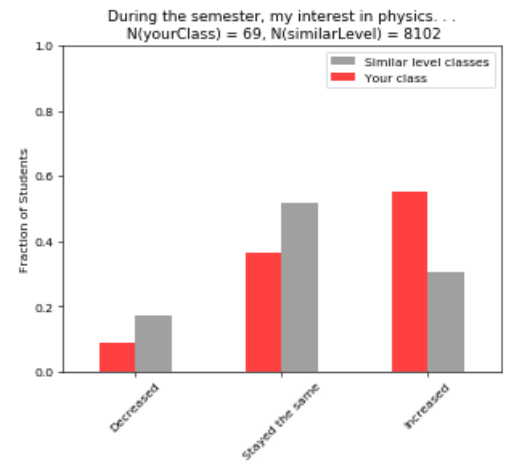}\hspace{5mm}\includegraphics[height=0.25\textheight]{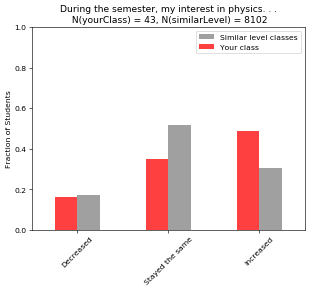}
\caption{E-CLASS results indicating the change in physics interest are shown. Students' interest in physics shows a significant increase in both the fall (left) and spring semesters (right). The interest increased slightly less in spring than it did in the fall.}
\label{fig:interestcompare}
\end{figure}

To compare the in-person labs to the remote labs, we draw from all of the instruments. We use the E-CLASS survey first to get a general sense of the impact of switching to remote instruction for a part of the semester. The overall E-CLASS score was reasonably close to the typical score in such classes. A comparison of the scores, before and after the semester, is shown in Figure \ref{fig:overallcompare}. In order to compare with an otherwise usual semester with all in-person labs, we also show the overall score for the same cohort of students from fall 2019. We see that despite the abrupt shift to remote instruction, the overall score for the spring semester remained nearly the same. Figure \ref{fig:interestcompare} shows the change in physics interest after the semester. We see a similar behaviour as fall, physics interest increased for a large number of students, more so than in a typical course. This increase is higher than typical classes at the same level. However, the proportion of the class for which interest increases is slightly smaller than in the fall. This could be indicative of the fact that the remote labs had to be prepared on a short notice and the students might have taken some time to adjust to the new paradigm, yet not having an adverse impact on the overall scores. 

\begin{figure}[h]
\centering
\includegraphics[height=0.25\textheight]{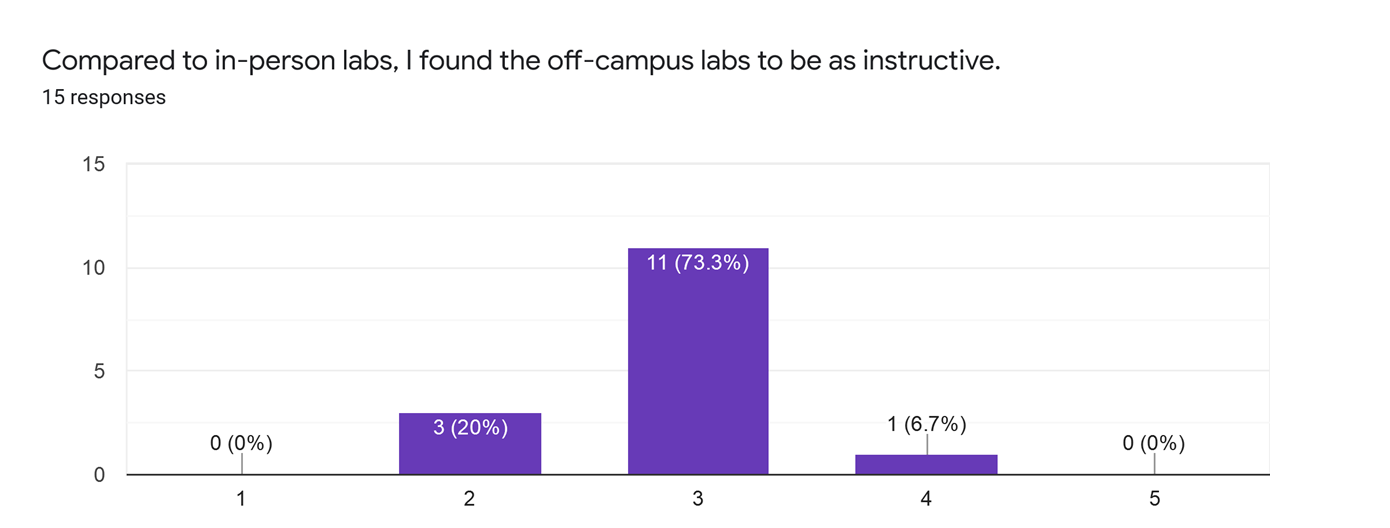}\hspace{5mm}\includegraphics[height=0.25\textheight]{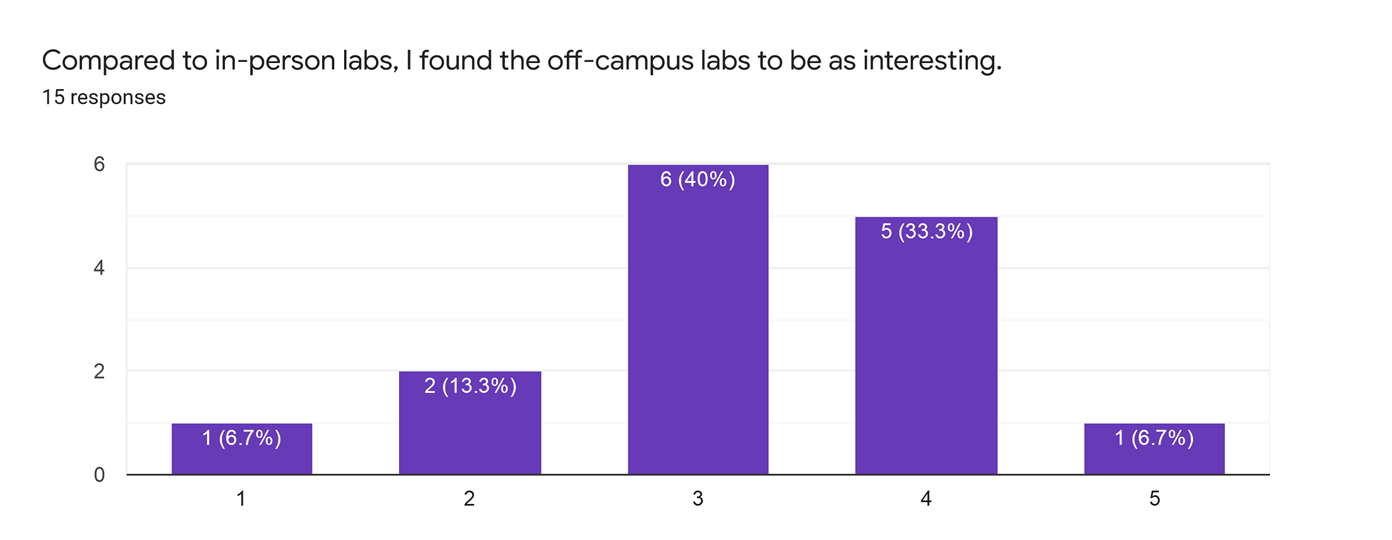}
\caption{Responses to an optional instructor created survey to assess how instructive and interesting the off-campus labs were when compared to in-person labs. Student responses to the two questions asked on the instructor-created survey. For both questions 1 corresponds to ``Far less instructive/interesting'' and 5 corresponds to ``Far more instructive/interesting''. The upper panel indicates that students found remote labs to be roughly as instructive as in-person labs. The lower panel indicates that students found remote labs to be slightly more interesting than in-person labs.}
\label{fig:OnOffCompare}
\end{figure}

To see the differences between in-person labs and remote labs for the spring semester more explicitly, we consider the results from the instructor-created survey, where students weigh the two paradigms based on interest and instructional quality. Students were asked ``Compared to in-person labs, I found the off-campus labs to be as instructive'' and ``Compared to in-person labs, I found the off-campus labs to be as interesting''. Responses to both questions were given on a Likert scale. The scale of the first question ranged from 1 = ``Far less instructive'' to 5 = ``Far more instructive''. The scale of the second question ranged from 1 = ``Far less interesting'' to 5 = ``Far more interesting''. The results of both questions can be seen in Figure \ref{fig:OnOffCompare}.

The interviews give us some additional information for this comparison. Each student was asked ``Compared to in-person labs, how `real' would you say this project was?''. The responses indicate that students think in-person labs feel real because of the professional equipment, but the process of the remote project lab felt more real. Below are some particularly interesting responses to this question.
    \begin{itemize}
        \item ``In in-person labs students were very reliant on TAs. If they had a hurdle, they would just approach the TAs. With the AHP project, they didn’t have the luxury of easily talking to TAs. It put greater focus on the student doing work independently, which is a goal of lab work.''
        \item ``The project was more scientific because it was so free form and students had to define the question.''
        \item ``In the lab things felt more real, because the equipment was more professional. But overall the Project felt like more scientific.''
        \item ``Being on campus, because it was more organized. Because there you were following a set of predefined steps. And there you have the right tools for the job, you don’t have to search for them.''
    \end{itemize}

Pulling all this information together, it appears that our remote labs were at least as effective as our in-person labs. Additionally, it seems that students felt the remote project lab was better than in-person labs at replicating the process of experimentation while also being slightly more interesting. The statistics of our study are such that no strong conclusion can be drawn, but it does seem promising that our transition to remote learning left us with labs that were \emph{at least} as effective as our in-person labs. We feel this is a significant result given the hasty nature in which we had to migrate to remote learning. 

\subsection{Comparing different remote labs}
Perhaps the most interesting thing to investigate in this study is how different types of remote labs compare to one another. For this comparison we use the instructor-created survey and the interviews, with the latter providing significant information. 

\begin{figure}[h]
\centering
\includegraphics[height=0.2\textheight]{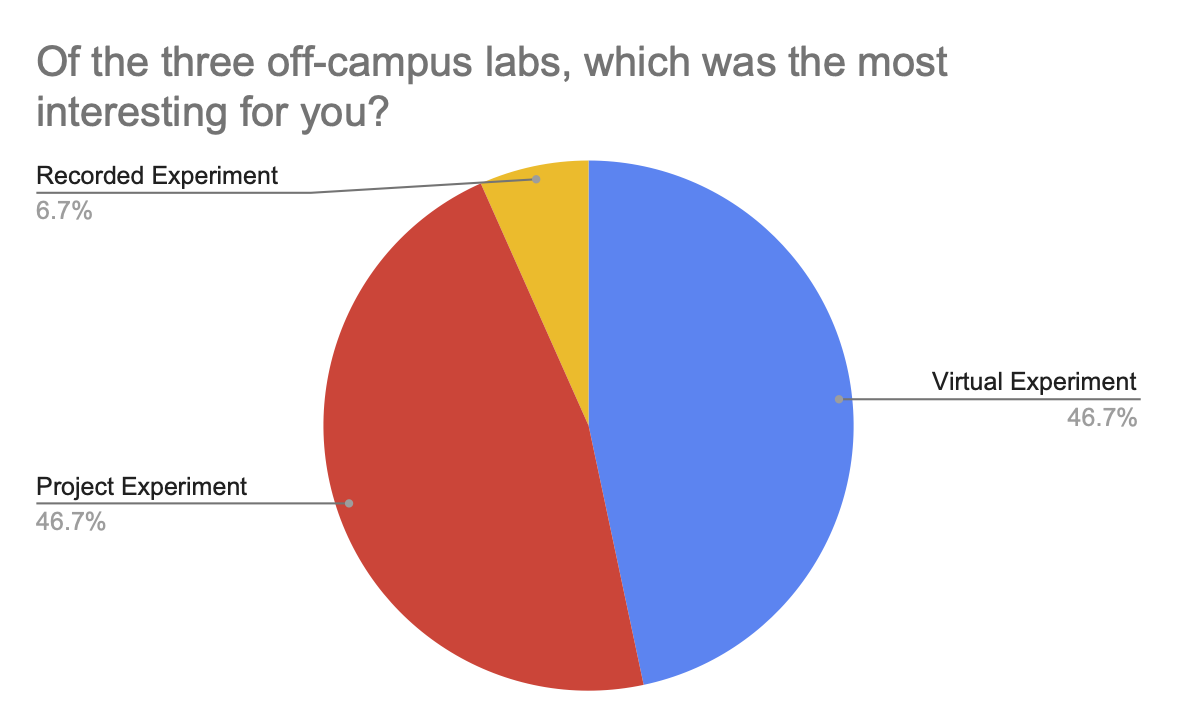}\hspace{5mm}\includegraphics[height=0.2\textheight]{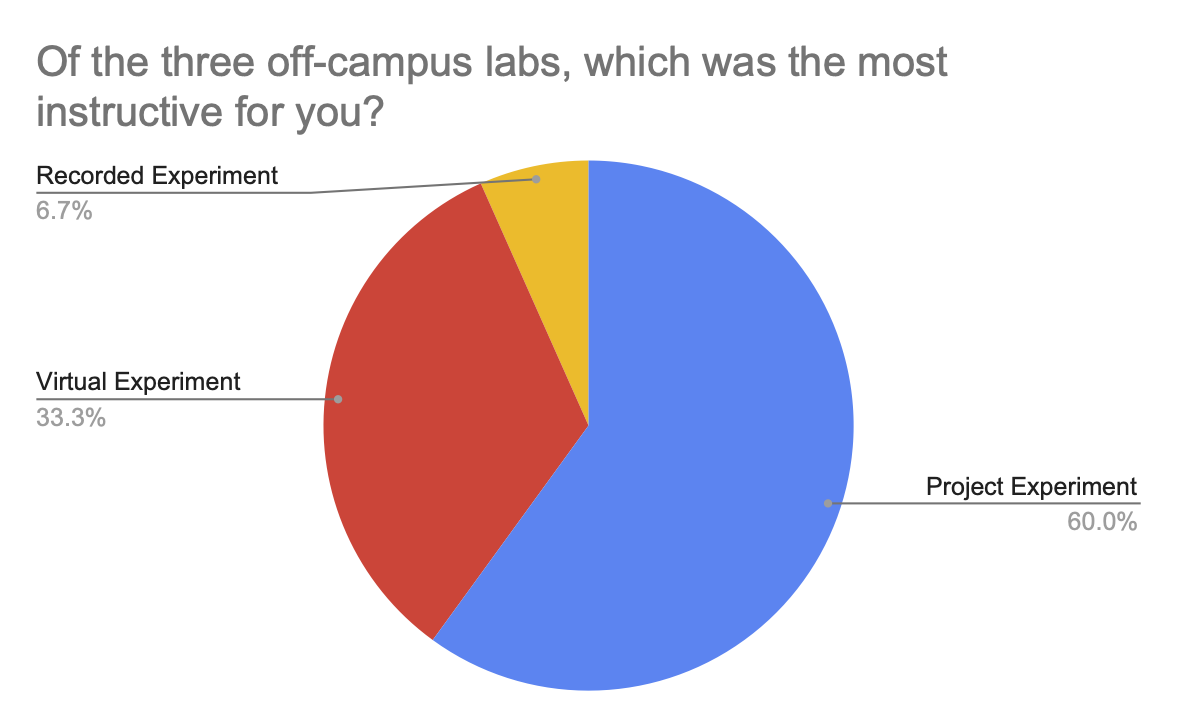}
\caption{The left panel shows the remote lab which students found the most interesting. The right panel shows the remote lab which students found the most instructive.}
\label{fig:OffCompare}
\end{figure}

The instructor-created survey asked two questions which directly compare the different types of remote labs: ``Of the three off-campus labs, which was the most instructive for you?''; ``Of the three off-campus labs, which was the most interesting for you?''. The results of these two questions can be seen in Figure \ref{fig:OffCompare}. The responses to these questions seem to indicate that students found the project lab and virtual experiment equally interesting, while finding the project lab more instructive. It is clear from the responses to both questions that few students found the recorded experiment to be the most instructive or interesting. These results are supported by the responses to interview questions, which are described below. 

During the interview, students were asked a series of questions that directly compare the three remote labs they completed. Before being asked the questions they were given the instructions ``For all the questions, keep in mind the goals of the labs are to make the student think critically about how to do experiments, and to teach experimental and data analysis skills.'' The goal of the questioning was to find out which labs students enjoyed the most and in which they thought they learned the most. 

The first relevant question was ``What was your favorite of the three labs? Why? How do these compare to the on-campus labs?''. The responses indicate a roughly equal split between virtual and project labs. The students that enjoyed the virtual experiment labs indicated this was because the were easier and provided a clearer understanding of the content. The students that enjoyed the project labs indicated this was because the process felt more like experimentation in a research lab. So while half of the students enjoyed the virtual experiment, their reasoning didn't align with the lab goals. Below are a few particularly interesting responses.  

        \begin{itemize}
            \item ``The virtual experiment was neat, and easy at a time when I needed it, but I didn’t feel like I learned that much. The project lab was the most interesting in the sense that I got to choose something I am interested in,  that was fun.”
            \item ``Definitely the project was the best of the three. It was far and away my favorite, even compared to on-campus. It felt more like labs I am accustomed to in research work in biology labs.''
            \item ``Project, virtual, recorded (most to least favorite). Compared to on campus, the virtual was good because you could get a good understanding of the concepts by manipulating parameters. The project shifted thinking toward what a real physicist would do, even more so than on-campus lab.''
            \item ``The virtual experiment was my favorite, because the way the questions flowed and were easy to understand. The way it was setup really helped me learn the concepts.'' 
        \end{itemize}
   
The second question relevant here was ```In which lab do you think you learned the most? Why?''. All but one student indicated that they learned the most in the project lab, and that student's response was ``The virtual experiment. For the project we picked a topic we knew already, so I didn’t learn concepts. In the project I learned the most about experimentation.'' The students that said they learned the most in the project lab indicated that this was because it made them understand all details of the lab, and be involved in every step. Below are a few particularly interesting responses. 
        \begin{itemize}
            \item ``The project lab, it made me think about what I needed, why, and how. It also forced me to come up with a question to test, and that was a new thing.''
            \item ``The project because I had to be involved in every step, and I really had to understand every single thing.''
            \item ``The project lab, there I learned mostly process, rather than content. It’s harder to learn content, but easier to learn process. I think there is potential with recorded experiment, but if you don’t understand what’s going on, you totally miss everything.''
            \item ``The project because it allowed me to get a hands-on understanding that I couldn’t get from the other two.''
        \end{itemize}

The final question relevant to this discussion was ``Given these goals, what type of labs would you assign?'' For this 75\% of the students said the project lab and the remaining 25\% said a mix of the project and the virtual experiment lab. Below are two particularly interesting responses. 
        \begin{itemize}
        \item ``Watching videos wasn’t as helpful as physically manipulating stuff. The project lab was my favorite, and the most fun I have ever had doing an experiment.''
        \item ``A mix of the virtual experiment and at home physics. There’s a lot to learn from having to come up with an experiment yourself. It forces you to learn a lot more. I had do a lot of reading up and learning. Then I had to get my data, see what it says, and see how I would present it. The at home physics was more rigorous. The virtual experiment was a bit easier, so a mix would be nice.'' 
        \end{itemize} 

In addition to responses from these questions, a majority of the students interviewed (75\%) indicated that that they found the process of the project lab the most interesting, saying things like ``The AHP was super fun. I had to figure how I was going to do everything, and I did a lot of trouble shooting. It was way more involved, which forced me to focus, and learn." and ``The project tasks were delegated and it was interesting to find a way to work as a team. And then the experimenter had to figure out everything on their own. It was time consuming, but the process of how to figure out how to optimize things was great. It was even more interesting than on-campus labs. It was very gratifying to see it finally work out.''

\begin{figure}[h]
\centering
\includegraphics[height=0.25\textheight]{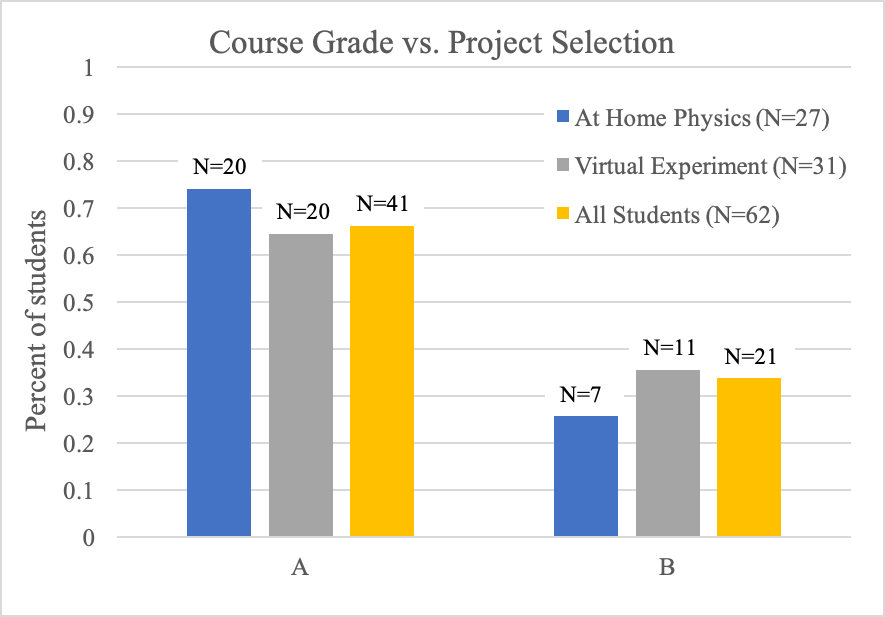}
\caption{A comparison between project selection and course grade is shown. This indicates no significant correlation between high-achieving students and project selection. Note that only 4 students did Algodoo projects and the grades of those students are not displayed; 1 of these students received an A and 3 received a B.}
\label{fig:GradeCompare}
\end{figure}

As previously mentioned, students were given three options for the project lab. We were interested in knowing if students of different achievement levels would choose different types of projects. For instance, do the highly motivated (and presumably highly achieving) students chose the potentially more involved At Home Physics project? Or do the students who are primarily interested in good grades choose the Virtual Experiment because it was perceived to be easier? To get at this, we investigated the student course grades as a function of project choices. The results are shown in Figure \ref{fig:GradeCompare}. There doesn't seem to be a significant correlation between project selection and course grade (note that students doing Algodoo projects are not displayed, as there were only four such students). 

\section{Conclusions}
To summarize, we have used three different instruments to assess the efficacy of remote labs in the introductory physics course at Princeton University (with 62 students), introduced halfway in the spring 2020 semester due to the COVID-19 pandemic. The three different types of labs studied, besides the in-person labs, were, a recorded experiment, a student defined three week project experiment, and an applet-based virtual experiment. We find that the overall learning attitude determined by the E-CLASS survey remains almost the same after the semester as before, corroborated by an instructor created survey, which reveals that the remote labs were perceived more interesting than in-person labs on an average. The same survey also brings out the distinction between the three remote labs, indicating that the project experiment and the virtual experiment generated more interest than the recorded experiments, while the project experiments were considered the most instructive. Interviews with 8 representative students explored the origins of these differences and a consensus emerged about project labs being perceived as an ideal experiment, wherein the students learned the most, while the virtual experiment was popular due to its ease and a well defined structure. These results are suggestive of an ideal remote lab combining the exploration of realistic physics experiments defined by the students themselves, to provide a creative and instructive approach, along with the ease and structure of a virtual applet based experiment.

\section{Future Work}
Based on this study, our fall labs will be a series of At Home Physics experiments. They will be scaffolded to slowly introduce the various aspects of designing and executing an experiment. Additionally, given their efficacy at teaching concepts, we plan to incorporate virtual experiments into the class portion of our course.\\ 
A full semester dedicated to AHP labs will provide us a better environment to assess their efficacy. Moreover, we will be able to directly compare them to our in-person labs. Our assessment will be similar to that followed here. Additionally we will implement the Physics Lab Inventory of Critical thinking \cite{PLIC2015} in order to get an independent assessment of how well our labs do at making our students think critically.
\bibliography{teaching}

\newpage
\appendix
\section{Rubrics}\label{app:Rubrics}
\includegraphics[width=.9\textwidth]{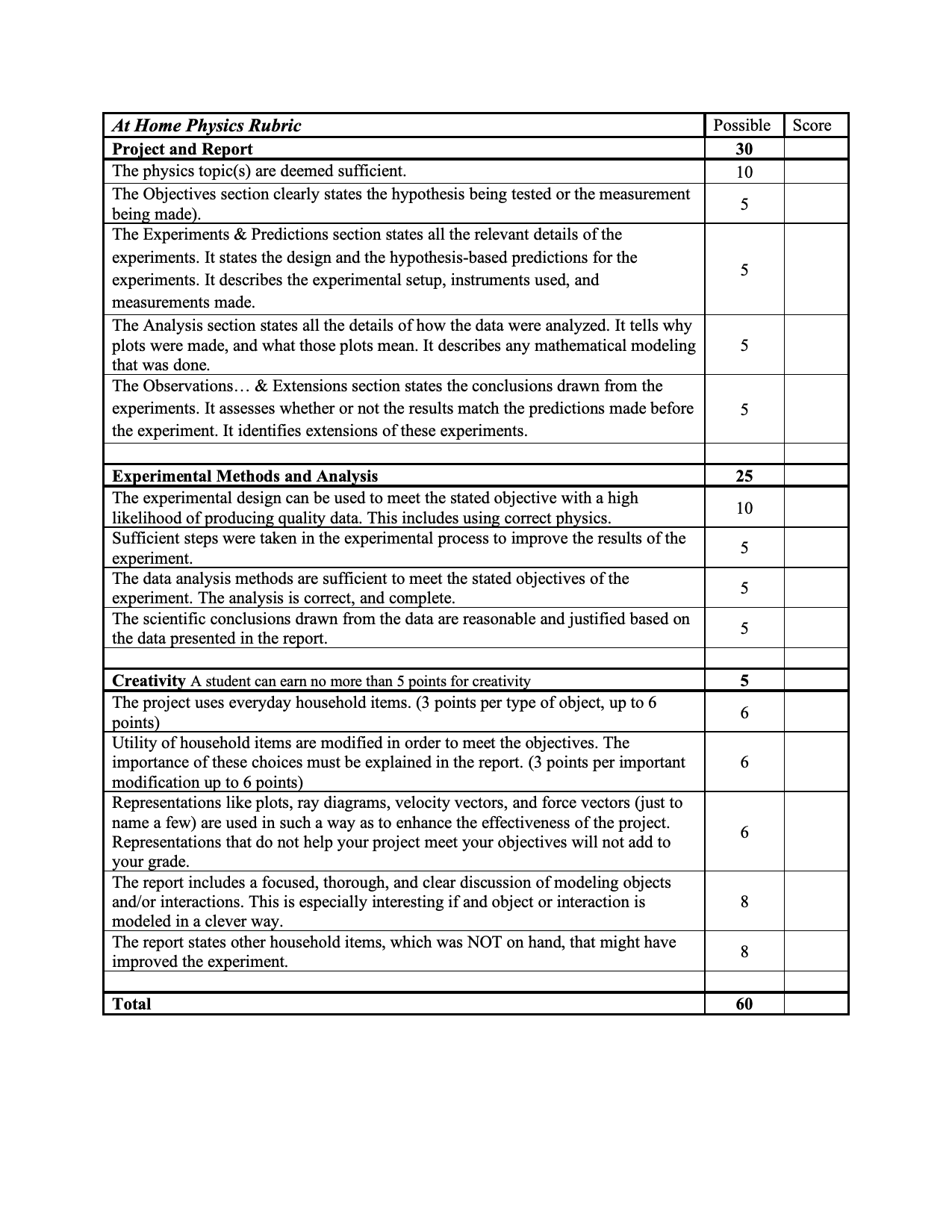}
\newpage
\includegraphics[width=.9\textwidth]{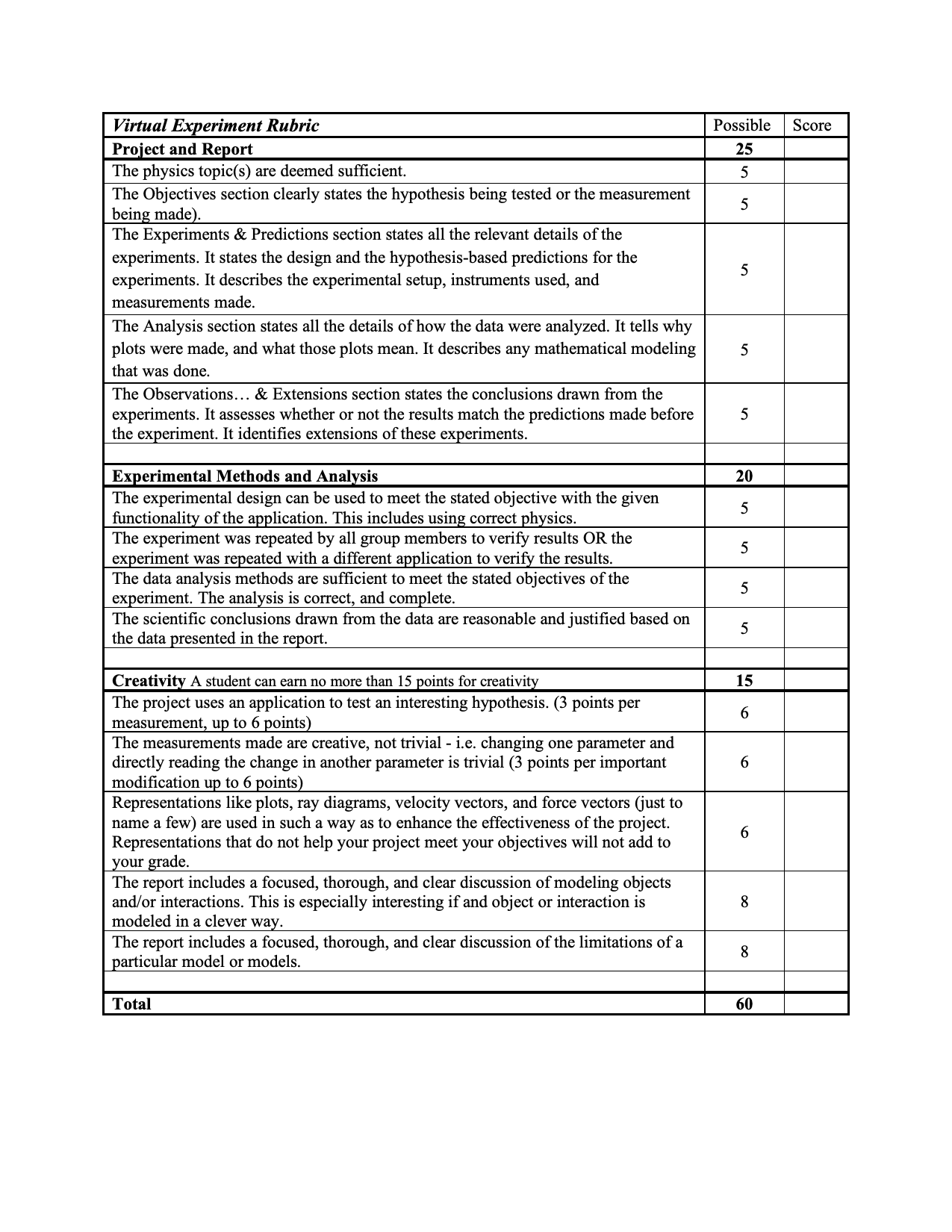}
\newpage
\includegraphics[width=.9\textwidth]{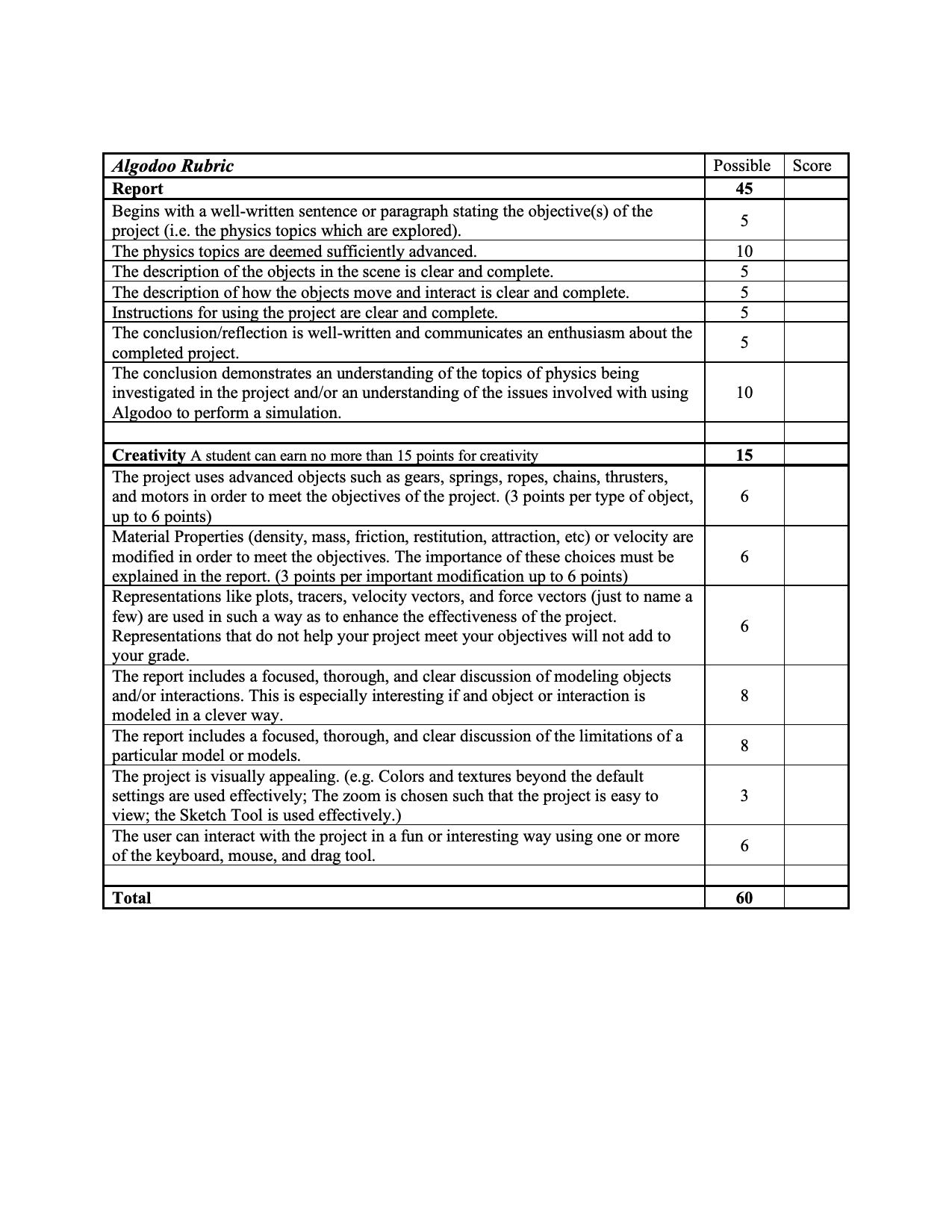}

\newpage

\section{Interview Questions}\label{app:Interview}
Questions comparing three different types of labs. 
\begin{itemize}
    \item For which of the three labs [Recorded, Project, Virtual] did you find the content the most interesting?
    \item For which of the three labs  [Recorded, Project, Virtual] did you find the process the most interesting?
    \item Which of the labs [Recorded, Project, Virtual] felt the most natural to you? That is, which do you think most closely matched your concept of a real physics experiment?
    \item Which type of lab [Recorded, Project, Virtual] was the most challenging for you? Why?
    \item What was your favorite of the three labs? Why? How do these compare to the usual labs?
    \item In which lab do you think you learned the most? Why? 
Given these goals, what type of labs would you assign? 
\end{itemize}

\noindent Questions specific to project lab. 
\begin{itemize}
    \item Why did you choose that type? Did you explore the other two projects before deciding or you were certain about the type from the start? Did the example projects for each type convey the essence of the process?
    \item If you had all the resources needed to do any one - any measurement device or tools for At Home, an app that could examine any phenomena you were interested in, and a hypothetical familiarity with Algodoo - which one would you pick?
    \item For At Home Physics
        \begin{itemize}
            \item How did you decide on your experiment?
            \item What obstacles did you encounter in doing the experiment? (e.g. defining a question, finding equipment, etc.)
            \item How many members in your group could perform the experiment, depending on equipment availability?
            \item Compared to in-person labs, how “real” would you say this project was?
        \end{itemize}
    \item For Virtual Experiments
        \begin{itemize}
            \item How did you decide on your experiment? Did you make the final decision based on the scope of the app or your interested topic from beforehand?
            \item What obstacles did you encounter in doing the experiment?
            \item What made this feel like an experiment, as opposed to a demonstration?
        \end{itemize}
\end{itemize}

\end{document}